# HARDWARE VIRTUALIZATION SUPPORT IN INTEL, AMD AND IBM POWER PROCESSORS

Kamanashis Biswas
Computer Science and Engineering Department
Daffodil International University
102, Shukrabad, Dhaka-1207, Bangladesh
ananda@daffodilvarsity.edu.bd

Md. Ashraful Islam
Department of Business Administration
Bangladesh Islami University
Gazaria Tower, 89/12, R. K. Mission Road, Dhaka-1203
ashraful47@yahoo.com

*ABSTRACT* – *At present, the mostly used and developed mechanism is hardware virtualization which provides a common platform to run multiple operating systems and applications in independent partitions. More precisely, it is all about resource virtualization as the term 'hardware virtualization' is emphasized. In this paper, the aim is to find out the advantages and limitations of current virtualization techniques, analyze their cost and performance and also depict which forthcoming hardware virtualization techniques will able to provide efficient solutions for multiprocessor operating systems. This is done by making a methodical literature survey and statistical analysis of the benchmark reports provided by SPEC (Standard Performance Evaluation Corporation) and TPC (Transaction processing Performance Council). Finally, this paper presents the current aspects of hardware virtualization which will help the IT managers of the large organizations to take effective decision while choosing server with virtualization support. Again, the future works described in section 4 of this paper focuses on some real-world challenges such as abstraction of multiple servers, language level virtualization, pre-virtualization etc. which may be point of great interest for the researchers.*

## 1. INTRODUCTION

A current trend in the computer industry is replacing uniprocessor computers with small multiprocessors [11]. Traditionally, most small multiprocessors have been SMPs (Symmetric Multiprocessors) with two or more processor chips where each processor has equal access to memory and hardware devices. But now, the scenario is going to be changed and the manufacturers are trying to increase PC manageability, user productivity and so on. Many techniques are already working to support multiprocessor operating systems such as giant locking, asymmetric approaches, virtualization, K42 etc.

There are two approaches which are used for parallelized processors. First, Symmetric multithreading (SMT) [3] where two or more concurrently running program threads share processor resources, e.g. Intel Pentium 4 and Xenon processor [12], and the 2-way multithreaded Sony/IBM Cell processor . Second one is chip multiprocessors (CMPs) [5], which partitions the chip area into two or more mostly independent processor cores, e.g. IBM POWER4 architecture was released as a dual-core chip in 2001 [8].

However, to implement multiprocessor operating systems and providing dynamic environment many technologies are evolved. But the most common and continuously updated technology is virtualization as all the companies like Intel, AMD, IBM always keep focus on this area by developing new and new virtualization techniques. Generally, virtualization is the faithful reproduction of an entire architecture in software which provides the illusion of a real machine to all software running above it [10]. Precisely, virtualization is a framework or methodology of dividing the resources of a computer into multiple execution environments, by applying one or more concepts or technologies such as hardware and software partitioning, time-sharing, partial or complete machine simulation, emulation, quality of service, and many others. This can be applied by either software or hardware or both and also for Desktop computer as well as for the Server machine.

In software-only virtualization technique, a Virtual Machine Monitor (VMM) program is used to distribute resources to the current multiple threads. But this software-only virtualization solution has some limitations. One is allocation of memory space by guest operating systems where applications would conventionally run. Another problem is binary translation, i.e. the necessity of extra layer of communication for binary translation, in order to emulate the hardware environment by providing interfaces to physical resources such as processors, memory, storage, graphics cards, and network adapters [16]. So hardware virtualization technique is a good solution to face the above problems which works in cooperation with VMM. This virtualization technique provides a new architecture upon which the operating system can run directly, it removes the need for binary translation. Thus, increased performance and supportability ensured. It also enhances the reliability, supportability, security, and flexibility of virtualization solutions. So the keen interest is on hardware virtualization.

This paper focuses on the virtualization supports of current microprocessors and makes a comparison among various hardware virtualization techniques offered by various companies. As there are many companies in the market competing with their latest technologies and improved facilities so it is important to have a good understanding about the mechanisms they are using. However, hardware virtualization is raising its acceptability over other virtualization techniques as it provides transparency, legacy support, simplicity, monitoring facility and security which is the point of interest for industrial computing systems.

## II. DIFFERENT VIRTUALIZATION TECHNIQUES

In uniprocessor system, it often assumes only one process in the kernel. As a result, it simplifies the kernel instructions and cross-process lock is not required. But the scenario is changed when multiple processors execute in the kernel. That means adding SMP support changes the original operating system. Hence mechanisms for supporting multiprocessors operating systems are required. There are different ways of organizing a multiprocessor operating system such as giant locking, coarse-grained locking, fine-grained locking, asymmetric approaches, virtualization and API/ABI compatibility and reimplementation. But the virtualization technique is the important one as the developers are continuously upgrading this technology. At first, we describe software-only virtualization and hardware virtualization. Then paravirtualization and full virtualization is explained.

### A. SOFTWARE ONLY VIRTUALIZATION

In software-only virtualization technique, the concept of 2-bit privilege level is used: using 0 for most privileged software and 3 for least privileged those. In this architecture (IA-32 and Itanium), the guest operating systems each communicates with the hardware through the Virtual Machine Monitor (VMM) which must decide that access for all virtual machines on the system. Thus, the virtual machine can be run on non-privileged mode i.e. non-privileged instructions can be executed directly without involving the VMM. But there are some problems that arise in software-only solution. Firstly, ring aliasing- problems that arise when software is run at a privilege level other than the level for which it was written. Secondly, address-space compression- occurs when guest software tries to access the VMM's guest's virtual address space. Thirdly, impacts on guest transitions- may cause a transition to the VMM and not to the guest operating system. VMMs also face other technical challenges such as use of private memory for VMM use only, use of VMM interrupt handling, hidden state access etc. [16].

### B. HARDWARE VIRTUALIZATION

Hardware virtualization allows the VMM to run virtual machines in an isolated and protected environment. It is also transparent to the software running in the virtual machine, which thinks that it is in exclusive control of the hardware. In 1999, VMware introduced the hosted VMM, and it was capable of extending a modern operating system to support a virtual machine that acts and runs like the hardware level VMM of old [14]. To address the problems of software only virtualization solution, hardware virtualization mechanism is applied which is possibly the most commonly known technology, including products from VMware and Microsoft's Virtual Server. Now, VMMs could run off-the-shelf operating systems and applications without recourse to binary translation or paravirtualization. This capability greatly facilitates the deployment of VMMs and provides greater reliability and manageability of guest operating systems and applications.

### C. PARAVIRTUALIZATION

Basically, to overcome the virtualization challenges of software-only virtualization, the VMM was developed by the designers that modify guest software (source or binary). Denali and Xen are examples of VMMs that use source level modifications in a technique called paravirtualization. Paravirtualization is similar to hardware emulation because in concept it is designed to support multiple OSs. The only implementation of this technology today is the Xen open source project, soon to be followed by an actual product from XenSource. Paravirtualization provides high performance and eliminates the 'changes to guest applications'. But the disadvantage is that it supports limited numbers of operating systems. For example, Xen cannot support an operating system that its developers have not modified, such as Microsoft Windows.

### D. FULL VIRTUALIZATION

Full system virtualization provides a virtual replica of the system's hardware so that operating systems and software may run on the virtual hardware exactly as they would on the original hardware [13]. The first introduced software for full virtualization system was CP-67, designed as a specialized time-sharing system which exposed to each user a complete virtual System/360 computer. Though full virtualization on PC architectures is extremely complex, at present it is pioneered in the market since 1998 as VMware initiated x86 based virtualization providing the fundamental technology for all leading x86-based hardware suppliers. It creates a uniform hardware image that implemented through software on which both operating system and application programs can run.

## III. HARDWARE VIRTUALIZATION SUPPORT IN MICROPROCESSORS

The challenges imposed on IT business that the CIOs and IT managers always face are cost-effective utilization of IT infrastructure and flexibility in adapting to organizational changes. Hence, virtualization is a fundamental technological innovation that provides the skilled IT professionals to organize creative solutions to those business challenges. The leading companies of IT sector are also introducing their innovative and well-developed approaches every day to cope with demands of the age. Again the hardware virtualization support is an important factor for the field of Grid Computing or secure on-Demand Cluster computing. The hardware support for virtualization in current microprocessors is addressed in this section.

## A. INTEL HARDWARE SUPPORT

Intel is developing microprocessors with various advanced virtualization supports. They are updating their technologies constantly to facilitate the users' demands. Starting with Server and mainframe systems virtualization, now Intel is providing hardware support for processor virtualization through virtual machine monitor software which is also known as hypervisor. The actual aim of using hypervisor is to arbitrate access to the underlying physical host system's resources so that multiple operating systems that are guests to VMM, can share them. IA-32 and Itanium architecture were built on software-only virtualization support [16]. But unfortunately they faced many challenges while providing virtualization supports. The software cannot work properly in concern with the core hardware, that's why it has to use complex schemes to imitate hardware features to the software. Moreover, it has to make the illusion that the host operating system thinking the virtual machine as another application. To eliminate these problems, VMM designers developed new solutions like Xen [2] and Denali VMMs that use source level modification known as paravirtualization. But the main limitation of this scheme is that it is applicable for a certain number of operating system. Hence Intel developed new architectures VT-x and VT-i for IA-32 processors (Core Duo and Solo) and Itanium processors family respectively which offered full virtualization using the hypervisor support. This new architecture enables VMM to run off-the-self operating systems and applications without any binary translation or paravirtualization. As a result it increases robustness, reliability and security.

## B. AMD HARDWARE SUPPORT

AMD has introduced their new Quad-Core AMD Opteron Processor (based on Pacifica specification) which is designed to provide optimal virtualization. This processor provides a number of features which enhances the performance and efficiency of the virtualization support. Firstly, AMD Opteron Rapid Virtualization Indexing, which allows virtual machine to more directly manage memory to improve performance on many virtualized applications [1]. It also decreases the "world-switch time" i.e. time spent switching from one virtual machine to another. Secondly, direct CPU-to-memory, CPU-to-I/O, and CPU-to-CPU connections to streamline server virtualization is ensured through AMD's direct connect architecture. Thus, it is possible to host more VMs per server and maximize the benefits of virtualization in terms of high bandwidth, low latency, and scalable access to memory. Thirdly, tagged Translation Look-Aside Buffer (TLB) has increased responsiveness in virtualized environments. Actually, through Tagged TLB, AMD Opteron processor maintains a mapping to the VMs individual memory spaces which eliminates additional memory management overhead and reduces switching time of virtual machines. Finally, Device Exclusion Vector (DEV) performs security checks in hardware rather than software. DEV mechanism controls access to virtual machine memory based on permission. These unique features have brought AMD to the frontline of battle on hardware virtualization support.

## C. IBM HARDWARE SUPPORT

As the successor of POWER3 and POWER4, IBM introduced advanced virtualization capabilities in IBM POWER5 processors in 2004. This processor includes increased performance and other functional enhancements of virtualization- reliability, availability, and serviceability in both hardware and software levels [9]. It uses hypervisor which is the basis of the IBM virtualization technologies on Powers systems. This technology provides fast page mover and simultaneous multithreading which finally extends the capability of PPC5. It supports logical partitioning and micro partitioning. Up to ten LPARs (logical partitions) can be created for each CPU. Thus the biggest 64-Way system is able to run 256 independent operating systems. Memory, CPU-Power and I/O can be dynamically controlled between partitions. Thus, IBM PPC5 uses the paravirtualization or cooperative partitioning in conjunction with the ATX, i5/OS, and Linux operating systems which offers minimal overhead [7]. This also ensures efficient resource utilization through recovery of idle processing cycles, dynamic reconfiguration of partition resources, and consolidation of multiple operating systems on a single platform and platform enforced security and isolation between partitions. The latest processor of IBM with virtualization support is IBM POWER6- the world's fastest computer chip, features industry leading virtualization capabilities. This processor provides a number of attractive features such as live partition mobility, expanded scalability, dynamic reallocation of resources etc. [6]. The Live Partition Mobility (LPM) feature allows clients to move running partitions automatically from one POWER6 server to another without powering down the server. Moreover, clients can create up to 160 virtual servers in a single box which provides much capability to run all kinds of different workloads (such as large scale database transactions to web servers) on the same server. IBM has built dynamic reallocation capabilities in chip. Users or in some cases the chip, itself, can reallocate and reassign computing resources within shared the environment. In addition to these exclusive features, IBM POWER6 provides enhanced performance, increased flexibility, application mobility etc.

## IV. FUTURE CHALLENGES AND SUPPORTS

Though hardware virtualization support in current processors has resolved many problems, it may also provide new solution for some future challenges. Extension of existing operating systems to present the abstraction of multiple servers is required for implementation of virtualization at other levels of the software stack. Language level virtualization technologies may be introduced by the companies to provide language run-times that interpret and translate binaries compiled for abstract architectures enable portability. Today Sun's Java and Microsoft's CLR VMMs dominate the market for language level virtualization technologies [4]. Memory virtualization should be efficient enough to make frequent changes to their page tables. Moreover, research must look at the entire data center level and surely significant strides will be made in this area in the incoming decade. At present, the manual migration is on practice, but the future should launch a virtual machine infra-structure that automatically performs load balancing, detects impending hardware failures and migrates virtual machines accordingly and creates and destroys virtual machines according to demand for particular support. To facilitate these supports, the instruction sets should be changed or new instructions should be added on which the processors can perform its jobs. Moreover, besides full and paravirtualization, pre-virtualization is a new technique that claims that it eliminates the guest-side engineering cost, yet matches the runtime performance of paravirtualization [17]. The virtualization will be useful for the industrial support, precisely for providing Grid services. At present, grid computing is gained prominence for its usage in physics, mathematics and medicine, to name just a few applications [10]. So grid computing or on-demand computing requires the virtualization support for taking the advantages of this technology in their shared computing environment.

## V. DISCUSSION

This paper describes an important part of the current computing systems as the present trend is to provide multiprocessor operating systems. The main objective of our study is to find out the current scenario of the hardware virtualization supports provided by various companies. From our survey, it is found that IBM is providing the best hardware virtualization supports where high availability, optimum system performance and efficiency are ensured. The most important feature incorporated in IBM technology is that the users have much control on shared resources as it is possible to modify the memory and I/O configurations in real time, without downtime, by the POWER6 server clients [6]. On the other hand, when IBM emphasizes on load balancing and live partition mobility, AMD focuses on intercommunication speed and performance such as high-bandwidth, low-latency access to memory, high throughput responsiveness for applications etc. Featuring AMD Virtualization technology with Rapid Virtualization Indexing and Direct Connect Architecture, Quad-Core AMD Opteron processors enable industry leading virtualization platform efficiency [1]. Intel, the giant microprocessor manufacturer improves the existing software-only virtualization solutions by enhancing the reliability, supportability, security and flexibility of virtualization solutions. Intel is working on increased hardware virtualization supports for both server and desktop computer. However, the table 5.1 shown in the next page addresses some important features of Intel, AMD and IBM processor with virtualization support.

From the table, it is clear that IBM POWER6 is the most powerful machine with enhanced virtualization capabilities. The fastest microprocessor used in POWER6 has hit speed of 6Hz, for the first time ever. Although IBM offers better output as it uses robust hardware support for virtualization, it is more costly than Intel and AMD. The user interaction also makes a security hole and vulnerable to the intruders. The virtualization technologies offered by Intel and AMD are not compatible, but each offers similar functionalities. These virtualization-friendly extensions for the x86 architecture essentially provide the foundation to maximize the efficiency and capabilities of software virtualization.

All microprocessor manufacturer companies are interested to enhance their virtualization capabilities. The main reason behind that hardware virtualization reduces the cost and provides reliability, availability & scalability. However, the concentration is more on server than the desktop computers as server machine requires more processing capacities. The table 5.2 in the next page shows the TPC Benchmark results (www.tpc.org) of AMD, Intel and IBM system with two server processors.

This table indicates that IBM P6 provides the best performance among all as well as it results in highest price. IBM P6 can perform 404462 transactions per minute while it is 273666 for Intel. On the other hand, AMD is less than half of total transactions per minute performed by Intel. In terms of price, IBM P6 system costs four times more than Intel while AMD is almost same as Intel.

Finally, the question is which forthcoming technology is going to overwhelm the others. The answer is hardware assisted virtualization techniques will dominate over all. From Power 5, IBM provides micro-partitioning and special technology for dynamic resource allocation. Again, AMD Opteron introduces tagged TLB, and Direct Connect Architecture which is designed for dedicated memory access and efficient switching between virtual machines. The integrated memory controller of AMD Opteron also improves overall virtualization performance and efficiency. And the most exciting news is that Intel also switches to hardware assisted virtualization techniques from their Intel Quad Core Xeon processor (7400 Series) which includes Intel VT FlexPriority for interrupt handling and Virtual Machine Device Queues (VMDq) to off-load the network I/O management burden and freeing processor cycles and improving overall performance. So, there is no doubt that the next advancement of hardware virtualization technology will be fully based on hardware assisted techniques.

## VI. CONCLUDING REMARKS

Some problems that must be specified which we faced continuing the study. Firstly, to achieve the exact result it is required to get access in real hardware and it is not possible. Secondly, the SPEC results is not specific to virtualization support only, it includes the virtualization features within processors. So the performance measurement partly takes the virtualization support in consideration. But the good news is that Standard Performance Evaluation Corporation (SPEC) has created a working group to address the development of a set of industry standard methods to compare performance of virtualization technologies. Current members of the working group include AMD, Dell, Fujitsu Siemens, Hewlett-Packard, Intel, IBM, Sun Microsystems, SWsoft(Now Parallels) and VMware [15]. So, to draw a sound and more accurate conclusion we have to wait few more days. But this paper will definitely provide the basis to explore one's journey towards hardware virtualization.

Table 5.1: Comparative view of Intel, AMD and IBM virtualization support based on SPEC evaluation [15].

| Characteristics | Intel Xeon 7000 Series Processors | AMD Opteron Processors | IBM POWER6 Processors |
|---|---|---|---|
| Hardware Assisted Virtualization | Intel Virtual Technology (VT) | AMD-V with Rapid Virtualization Indexing | Live Partition Mobility (LPM) |
| Modular, Glueless, Scalability | Requires Northbridge | Yes | Yes, supports up to 160 virtual servers |
| SMP Capabilities | Up to 4 Sockets/ 16 Cores | Up to 8 Sockets/ 32 Cores | Up to 8 Sockets/ 16 Cores |
| User Interaction | No | No | Yes, user can create VMs which span the entire system |
| Server/ Application Downtime | Yes | Yes | No, Sustain system availability during maintenance or re-hosting |
| Concurrent firmware and Operating System Updates | No | No | Yes, even when applications are active |

Table 5.2: TPC Benchmark results with price and performance (based on SPEC)

| Spec Revision | tpmC (transaction per minute) | Price/Performance | Total System Cost (USD) | Server CPU Type | Total Server CPU's | Total Server Processors | Total Server Cores | Total Server Threads |
|---|---|---|---|---|---|---|---|---|
| 5.6 | 113628 | 2.99 | 338730 | AMD Opteron – (2.6 GHz) | 2 | 2 | 4 | 4 |
| 5.9 | 273666 | 1.38 | 376910 | Intel Quad-Core Xeon Processor X5460 (3.16GHz) | 2 | 2 | 8 | 8 |
| 5.8 | 404462 | 3.51 | 1417121 | IBM Power6 (4.7 GHz) | 2 | 2 | 4 | 8 |

______________________________________________________

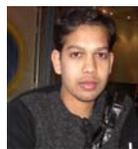
**Kamanashis Biswas**, born in 1982, post graduated from Blekinge Institute of Technology, Sweden in 2007. His field of specialization is on *Security Engineering*. At present, he is working as a *Senior Lecturer* in Daffodil International University, Bangladesh.

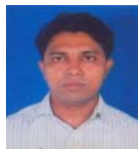
**Md. Ashraful Islam,** is post graduated from American Liberty University, UK. Now he is working as an *Assistant Professor* in Bangladesh Islami University. His major area of interest is *software engineering, e-learning and MIS*.